\documentclass{myaa}
\usepackage{graphicx}
\usepackage{txfonts}
\usepackage{balance}
\begin{document}

\title{
NLTE determination of the aluminium abundance in a homogeneous
sample of extremely metal-poor stars
\thanks{Based on observations obtained with the ESO Very Large
Telescope at Paranal Observatory, Chile (Large Programme ``First
Stars'', ID 165.N-0276(A); P.I.: R. Cayrel).}
}

\author {
S.M. Andrievsky\inst{1,2}\and
M. Spite\inst{1}\and
S.A. Korotin\inst{2}\and
F. Spite\inst{1}\and
P. Bonifacio\inst{1,3,4}\and
R. Cayrel\inst{1}\and
\\
V. Hill\inst{1}\and
P. Fran\c cois\inst{1}
}

\offprints{M. Spite\\
           e-mail: Monique.Spite@obspm.fr}
\institute {
   GEPI, CNRS UMR 8111, Observatoire de Paris-Meudon, F-92125 Meudon
   Cedex, France,
\and
   Department of Astronomy and Astronomical Observatory, Odessa
   National University, Shevchenko Park, 65014 Odessa, Ukraine.
\and
   CIFIST Marie Curie Excellence Team
\and
   Istituto Nazionale di Astrofisica - Osservatorio Astronomico di
   Trieste, Via Tiepolo 11, I-34143, Trieste, Italy 
}

\date{}

\authorrunning{Andrievsky et al.}
\titlerunning{NLTE determination of the aluminium abundance in EMP 
stars}

  \abstract
   {}
   {Aluminium is a key element to constrain the models of the chemical
   enrichment and the yields of the first supernovae. But obtaining
   precise Al abundances in extremely metal-poor (EMP) stars requires 
   that the non-LTE effects be carefully taken into account.}
   {The NLTE profiles of the blue resonance aluminium lines have been 
   computed in a sample of 53 extremely metal-poor stars 
   with a modified version of the program MULTI 
   applied to an atomic model of the Al atom with 78 levels of Al~I 
   and 13 levels of Al~II, and compared to the observations.
   }
   {With these new determinations, all the stars of the sample show a
   ratio Al/Fe close to the solar value: $\rm [Al/Fe] =-0.06 \pm0.10$
   with a very small scatter.  These results are compared to the
   models of the chemical evolution of the halo using different models
   of SN~II and are compatible with recent computations.
   The sodium-rich giants are not found to be also aluminium-rich and
   thus, as expected, the convection in these giants only brings to
   the surface the products of the Ne-Na cycle.  
   }
   {}

\keywords{ Line : Formation -- Line : Profiles -- Stars: Abundances --
Stars: Mixing -- Stars: Supernovae -- Galaxy evolution}

\maketitle

\section{Introduction}

In the  early stages of Galactic evolution, aluminium nuclei are built by
massive SN~II. Aluminium is produced by hydrostatic carbon and neon
burning and is later expelled in the interstellar medium during the
explosion.  The explosion does not affect the yields significantly.
At low metallicity the aluminium production is mainly based on $\rm
^{12}C$ produced by He burning: it can be considered as a primary
production (independent of the metallicity of the supernova).  However
it has been shown that the aluminium yields depend on the available
neutron excess $\rm \eta$ and thus on the metallicity of the
progenitor.  Aluminium can then behave as a secondary element (see
Arnett, 1971, 1996, and also Gehren et al., \cite {GSZ06}).  At low
metallicity it is expected that the primary production dominates and,
as a consequence, the ratio [Al/Fe] should be almost constant.

Available data on LTE aluminium abundances in stars of different
metallicities (see e.g. Figure 10 of Samland 1998, Baum\"uller \&
Gehren 1997 and references therein,
 Norris et al.  2001, Tsujimoto et al.
2002, Zhang \& Zhao, 2006) show that the behaviour of the ratio
[Al/Fe] is quite complex and 
the scatter at low metallicity is so large that it is difficult to
determine a trend.
Standard models of chemical evolution are not able to predict the
scatter of abundance ratios.  One mechanism for introducing the
required inhomogeneity is e.g. the model proposed by Tsujimoto et al.
(1999) in which each newly formed star inherits the elemental
abundance pattern of an individual SNe (for an   inhomogeneous model, see also
Argast et al.  2000).
With this hypothesis Tsujimoto et al.  (2002) could deduce from the
large scatter of the aluminium abundance at low metallicity, the yields
of aluminium in massive supernovae as a function of their metallicity
($\rm m_{Al} \propto Z^{0.6}$ for $\rm [Mg/H] < -1.8$).

This is possible only if we can assume that the aluminium abundance
measured in the atmosphere of the old stars reflects the abundance in
the interstellar medium at the time of the star formation.  For
example a mixing between the surface and the deep layers of the stars
where aluminium is formed (AGB stars), could artificially increase the
original abundance of aluminium in the atmosphere of the stars (Herwig
2005, Spite et al.  2006).

Furthermore, NLTE effects, generally neglected, can affect the
abundance computations as a function of the temperature, the gravity
and the metallicity of the star.  At low metallicity the NLTE
corrections become rather large, in particular because at $\rm
[Fe/H]<-2.5$ the aluminium abundance can be deduced only from the
resonance lines at 3944 and 3962 \AA\  which are known to be
strongly affected by NLTE effects.

In this paper we present a homogeneous NLTE determination of the
aluminium abundance in a sample of 51 normal (not carbon-rich)
Extremely Metal Poor (EMP) stars; 18 are turnoff stars, 33 giants and
among them 29 have $\rm[Fe/H]<-3$.  These stars have been 
analysed (LTE analysis) by Cayrel et al.  (\cite{CDS04}) and Bonifacio
et al.  (\cite{BMS07}).  In a previous paper (Andrievsky et al. 2007)
we have determined the sodium abundance in this sample of
stars taking into account NLTE effects.

Several  papers have taken into account NLTE line formation
in the determination of the aluminium abundance in metal-poor stars
(mainly Gehren et al.  2004, 2006) and the scatter of the ratios
[Al/Mg] or [Al/Fe], in the considered metallicity range,  has been
strongly reduced.  The sample of Gehren et al., does not reach a
[Fe/H] value lower than about --2.5;  our sample extends this work
down to [Fe/H]= --4.

\section {The star sample} 
The observed spectra of the stars investigated here have been
presented in detail in Cayrel et al.  (2004) and Bonifacio et al.
(\cite{BMS07}).   
However, in the present paper the carbon-rich stars of their sample, 2
giants and 1 turnoff star, that need specific models, have not been
taken into account and will be studied later.

The observations were performed with the high resolution spectrograph
UVES at ESO-VLT. The resolving power of the spectrograph in the region
of the blue aluminium lines is $R=47000$, the S/N per pixel in this
region of the spectra is close to 120 and there are about 5 pixels per
resolution element.

The fundamental parameters of the models (Teff, log g, metallicity)
have been derived by Cayrel et al.  (\cite{CDS04}) for the giants and
Bonifacio et al.  (\cite{BMS07}) for the turnoff stars.  Briefly,
temperatures of the giants are deduced from the colours with the
calibration of Alonso et al.  (\cite{AAM99}, \cite{AAM01}), and
temperatures of the turnoff stars from the wings of the $\rm
H{\alpha}$ line.

For the turnoff stars, the temperatures deduced
from the colours with the calibration of Alonso et al.  (\cite{AAM96})
are in good agreement with the $\rm H{\alpha}$ temperatures (standard
deviation 100\,K).  On the contrary, the temperatures derived from the
Ram\'{\i}rez \& Mel\'{e}ndez (\cite{RM05}) calibration are
considerably higher and incompatible with the $\rm H{\alpha}$
temperatures.  Moreover, with the calibration of Alonso et al.
(\cite{AAM96}) the abundance trend of the Fe~I lines with excitation
potential is less than 0.06dex/eV. On the contrary with the high $\rm
T_{eff}$ derived from the Ram\'{\i}rez and Mel\'{e}ndez calibration, no
iron-excitation equilibrium is achieved (see Bonifacio et al.
(\cite{BMS07}) for more details).

These discrepancies suggest that a systematic error in the adopted
temperatures of the order of 200\,K is possible (Bonifacio et al.,
\cite{BMS07}) although unlikely.

The gravities are from the ionisation equilibrium of iron (in the LTE
approximation) and we note that they could be affected by non-LTE
effects.

The parameters of the models are repeated in Table 1 for the
reader's convenience.  

\begin{table*}
\begin{center}    
\caption[]{
Model parameters and NLTE aluminium abundance in our sample of EMP
stars.  The solar abundance of aluminium has been taken from Grevesse
and Sauval (\cite{GS00}) as in Cayrel et al.  (\cite{CDS04}): $\rm log
\epsilon_{(Al)~\odot}=6.47$.
}
\label{tabstars}
\begin{tabular}{lcccccrrrrr}
\hline
star      & $T_{\rm eff},\,K$ & $\log~g$ & v$_{\rm t}$, km~s$^{-1}$ &
[Fe/H] & log $\epsilon$(Al) & [Al/H] & [Al/Fe] &Rem\\
\hline
~~~\\
{\bf turnoff stars} \\
\hline
BS~16023--046  & 6360 & 4.5 & 1.4 &  --2.97  & 3.20  &  --3.27 &  --0.30 & \\
BS~16968--061  & 6040 & 3.8 & 1.5 &  --3.05  & 3.30  &  --3.17 &  --0.12 & \\
BS~17570--063  & 6240 & 4.8 & 0.5 &  --2.92  & 3.25  &  --3.22 &  --0.30 & \\
CS~22177--009  & 6260 & 4.5 & 1.2 &  --3.10  & 3.28  &  --3.19 &  --0.09 & \\
CS~22888--031  & 6150 & 5.0 & 0.5 &  --3.28  & 3.10  &  --3.37 &  --0.09 & \\
CS~22948--093  & 6360 & 4.3 & 1.2 &  --3.43  & 3.00  &  --3.47 &  --0.04 & \\
CS~22953--037  & 6360 & 4.3 & 1.4 &  --2.89  & 3.50  &  --2.97 &  --0.08 & \\
CS~22965--054  & 6090 & 3.8 & 1.4 &  --3.04  & 3.40  &  --3.07 &  --0.03 & \\
CS~22966--011  & 6200 & 4.8 & 1.1 &  --3.07  & 3.25  &  --3.22 &  --0.15 & \\
CS~29499--060  & 6320 & 4.0 & 1.5 &  --2.70  & 3.50  &  --2.97 &  --0.27 & \\
CS~29506--007  & 6270 & 4.0 & 1.7 &  --2.91  & 3.48  &  --2.99 &  --0.08 & \\
CS~29506--090  & 6300 & 4.3 & 1.4 &  --2.83  & 3.55  &  --2.92 &  --0.09 & \\
CS~29518--020  & 6240 & 4.5 & 1.7 &  --2.77  &~~~--  &   ~~~-- &   ~~~-- & \\
CS~29518--043  & 6430 & 4.3 & 1.3 &  --3.24  & 3.05  &  --3.42 &  --0.18 & \\
CS~29527--015  & 6240 & 4.0 & 1.6 &  --3.55  & 2.97  &  --3.50 &    0.05 & \\
CS~30301--024  & 6330 & 4.0 & 1.6 &  --2.75  & 3.60  &  --2.87 &  --0.12 &\\
CS~30339--069  & 6240 & 4.0 & 1.3 &  --3.08  & 3.17  &  --3.30 &  --0.22 &\\
CS~31061--032  & 6410 & 4.3 & 1.4 &  --2.58  & 3.70  &  --2.77 &  --0.19 &\\
\hline
~~~\\
{\bf giants}  \\
\hline
HD~2796        & 4950 & 1.5 & 2.1 &  --2.47 &  4.08  & --2.39 &    0.08 & m\\
HD~122563      & 4600 & 1.1 & 2.0 &  --2.82 &  3.55  & --2.92 &  --0.10 & m\\
HD~186478      & 4700 & 1.3 & 2.0 &  --2.59 &  3.95  & --2.52 &    0.07 & m\\
BD~+17:3248    & 5250 & 1.4 & 1.5 &  --2.07 &  4.50  & --1.97 &    0.10 & m\\
BD~--18:5550   & 4750 & 1.4 & 1.8 &  --3.06 &  3.35  & --3.12 &  --0.06 & \\
CD~--38:245    & 4800 & 1.5 & 2.2 &  --4.19 &  2.35  & --4.12 &    0.07 & m\\
BS~16467--062  & 5200 & 2.5 & 1.6 &  --3.77 &  2.60  & --3.87 &  --0.10 & \\
BS~16477--003  & 4900 & 1.7 & 1.8 &  --3.36 &  3.00  & --3.47 &  --0.11 & \\
BS~17569--049  & 4700 & 1.2 & 1.9 &  --2.88 &  3.50  & --2.97 &  --0.09 & m\\
CS~22169--035  & 4700 & 1.2 & 2.2 &  --3.04 &  3.10  & --3.37 &  --0.33 & m\\
CS~22172--002  & 4800 & 1.3 & 2.2 &  --3.86 &  2.65  & --3.82 &    0.04 & \\
CS~22186--025  & 4900 & 1.5 & 2.0 &  --3.00 &  3.30  & --3.17 &  --0.17 & m\\
CS~22189--009  & 4900 & 1.7 & 1.9 &  --3.49 &  2.74  & --3.73 &  --0.24 & \\
CS~22873--055  & 4550 & 0.7 & 2.2 &  --2.99 &  3.45  & --3.02 &  --0.03 & m\\
CS~22873--166  & 4550 & 0.9 & 2.1 &  --2.97 &  3.45  & --3.02 &  --0.05 & m\\
CS~22878--101  & 4800 & 1.3 & 2.0 &  --3.25 &  3.10  & --3.37 &  --0.12 & m\\
CS~22885--096  & 5050 & 2.6 & 1.8 &  --3.78 &  2.80  & --3.67 &    0.11 & \\
CS~22891--209  & 4700 & 1.0 & 2.1 &  --3.29 &  3.15  & --3.32 &  --0.03 & m\\
CS~22896--154  & 5250 & 2.7 & 1.2 &  --2.69 &  3.84  & --2.63 &    0.06 & \\
CS~22897--008  & 4900 & 1.7 & 2.0 &  --3.41 &  2.95  & --3.52 &  --0.11 & \\
CS~22948--066  & 5100 & 1.8 & 2.0 &  --3.14 &  3.20  & --3.27 &  --0.13 & m\\
CS~22952--015  & 4800 & 1.3 & 2.1 &  --3.43 &  3.05  & --3.42 &    0.01 & m\\
CS~22953--003  & 5100 & 2.3 & 1.7 &  --2.84 &  3.45  & --3.02 &  --0.18 & \\
CS~22956--050  & 4900 & 1.7 & 1.8 &  --3.33 &  3.15  & --3.32 &    0.01 & \\
CS~22966--057  & 5300 & 2.2 & 1.4 &  --2.62 &  4.00  & --2.47 &    0.15 & \\
CS~22968--014  & 4850 & 1.7 & 1.9 &  --3.56 &  2.75  & --3.72 &  --0.16 & \\
CS~29491--053  & 4700 & 1.3 & 2.0 &  --3.04 &  3.36  & --3.11 &  --0.07 & m\\
CS~29495--041  & 4800 & 1.5 & 1.8 &  --2.82 &  3.60  & --2.87 &  --0.05 & \\
CS~29502--042  & 5100 & 2.5 & 1.5 &  --3.19 &  3.25  & --3.22 &  --0.03 & \\
CS~29516--024  & 4650 & 1.2 & 1.7 &  --3.06 &  3.15  & --3.32 &  --0.26 & \\
CS~29518--051  & 5200 & 2.6 & 1.4 &  --2.69 &  3.85  & --2.62 &    0.07 & m\\
CS~30325--094  & 4950 & 2.0 & 1.5 &  --3.30 &  3.25  & --3.22 &    0.08 & \\
CS~31082--001  & 4825 & 1.5 & 1.8 &  --2.91 &  3.55  & --2.92 &  --0.01 & \\
\hline
\end{tabular}
\end{center}
-A "m" in the last column means that the star is  "mixed" (see Spite 
et al., \cite{SCH06})\\
\end{table*}

\section {Determination of the aluminium abundance}
\subsection{Atomic model and NLTE calculations} \label{atomodel}

As discussed by Gehren et al.  (2004, and references therein), the
lines of aluminium are subject to a strong NLTE effect in
metal-poor stars because of the reduced role of the collisions, which are
responsible for thermalisation,
in their atmospheres (lower electron concentration) thus 
the populations of the aluminium
atomic levels are mainly controlled by radiative processes.

For our NLTE calculations we adopted an aluminium atomic model that
consists of 78 levels of Al~I and 13 levels of Al~II.
The energy levels are from Kaufman \& Martin (1991).  For all the
terms the fine structure was ignored, and they were treated as single
levels in the calculations.  However the ground level was treated as
two sublevels.

Radiative and collisional transitions were considered between the
first 45 Al~I levels ({\it n}~$<$~12, {\it l}~=~5) and the ground
level of Al~II. Transitions between the remaining levels were ignored,
and those levels were used in the calculations only to meet the
condition of the particle number conservation.  Oscillator strengths
and photoionisation cross-sections were taken from the TOPbase (see:
http://vizier.u-strasbg.fr/topbase/topbase.html).  Only for the levels
with $n = 10, ~l = 3-5$ was the photoionisation treated in the
hydrogen-like approximation.  Altogether, 288 bound-bound transitions
were considered in detail.

Electron impact ionisation was described by Seaton's (1962) formula,
while the electron impact excitation for the allowed bound-bound
transitions was estimated with  van Regemorter's (1962)
formula.  Collisional rates for the forbidden transitions were
calculated with a semi-empirical  formula provided by 
Allen (1973), with a
collisional force  equal to 1.  Inelastic collisions of aluminium
atoms with hydrogen atoms may play a significant role in the atmospheres
of cool stars.  We have taken into account this effect by using Drawin's
formula with a correction factor of 0.1 (see Steenbock \& Holweger
1992).

The NLTE aluminium abundance was determined with a modified version of
the MULTI code (Carlsson 1986).  All modifications are described in
Korotin et al.  (1999).  This modified version includes opacity
sources from ATLAS9 (Kurucz 1992); this enables one to perform
an accurate determination of the continuum opacity and intensity
distribution in the UV region, which is extremely important for the
correct determination of the radiative rates of the bound-bound
transitions in the aluminium atom.  Models of stellar atmospheres were
interpolated in Kurucz's grid with $\alpha = 1.25$.

After solving the coupled  radiative transfer and statistical
equilibrium equations, the averaged levels were split with respect to
their multiplet structure. Then the level populations were
redistributed proportionally to the statistical weights of the
corresponding sublevels, and, finally, the lines of interest were
studied.

\begin{figure}
\resizebox  {8.4cm}{4.2cm}
{\includegraphics {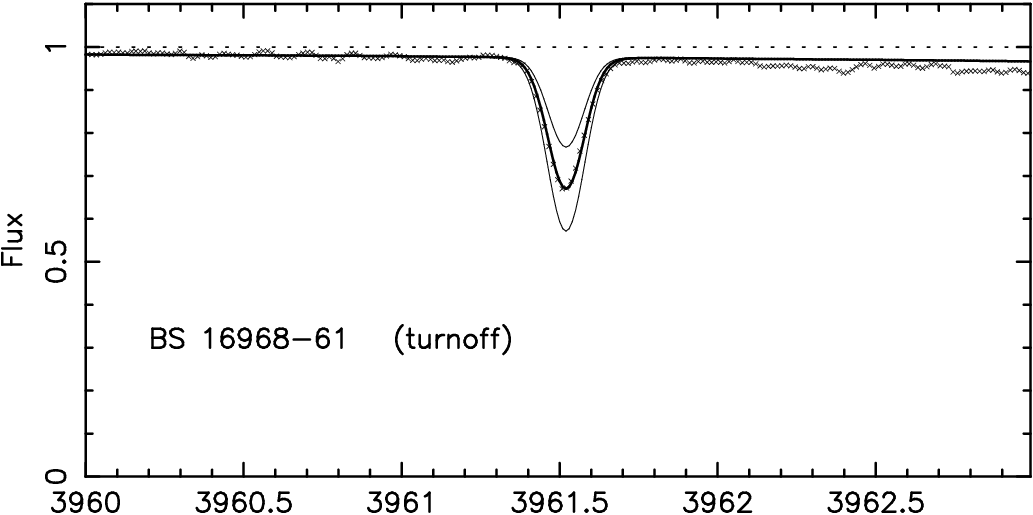} }
\resizebox  {8.4cm}{4.2cm}
{\includegraphics {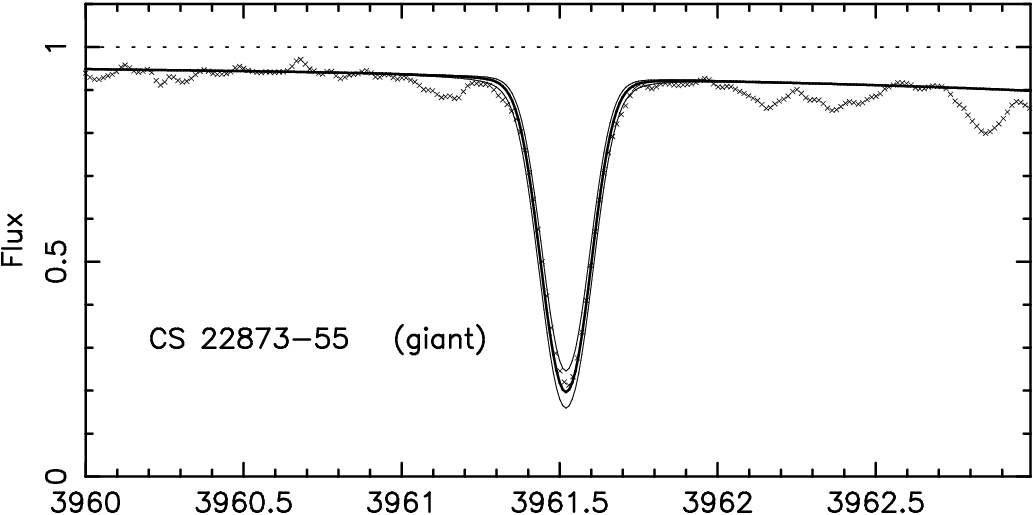} }
\caption[]{Profile fitting for a turn-off star (upper panel) and a 
giant star (lower panel). The best-fit Al abundance $\log 
\epsilon$(Al) was varied by $\pm 0.3$ dex.}
\label {spectra}
\end{figure}

In very metal-poor stars, only the UV resonance doublet of aluminium
is measurable.  For these lines the broadening parameters were taken
from the VALD data-base (http://ams.astro.univie.ac.at/vald/).  Van
der Waals $C_6$ parameter was taken from Gehren et al.  (2004).
Oscillator strengths of the studied lines are from Wiese \& Martin
(1980).

To verify the completeness of our adopted aluminium atomic model, we
have carried out the test calculations of the Al line profiles in the
solar spectrum.  The Solar Flux Atlas of Kurucz et al.  (1984) and
Kurucz's model of the solar atmosphere (1996) were used for this
purpose.  To take into account the chromospheric growth of the
temperature, this model was combined with a model of the solar
chromosphere from Maltby et al.(1986).  Not only were the profiles of the
resonance aluminium doublet calculated, but we have also analysed
the 6696-98\,\AA, 7835-36\,\AA, 8772-73\,\AA  lines.  The very good 
agreement between our calculated profiles and the observed
ones, for these aluminium lines gives an
independent confirmation that our adopted aluminium atomic model is
correct.  The solar aluminium abundance derived from these
computations was $ \rm log \epsilon_{(Al)}= 6.43$ for $ \rm log
\epsilon_{H}= 12.$

Since the resonance lines of aluminium are situated in the vicinity of
the $H$ and $K$ Ca~II lines and the H5 line, the wings of these lines have
to be properly computed.  Therefore, using the MULTI code we first
calculated the departure coefficients $b_{\rm i}$ for each atmosphere
model, and then these coefficients were used in synthetic spectrum
calculations that cover the whole region occupied by Ca~II and H5
lines (an
updated version of the
PC compatible synthetic spectrum code of Tsymbal 1996,  was used).

Two examples of the line profile fitting for the turn-off and giant
stars are shown in Fig.  \ref{spectra}.  This work is
mainly based on the line at 3961\,\AA; the line at 3944\,\AA\  is
often severely blended by CH lines.

The aluminium abundances $\rm log \epsilon_{Al}$, [Al/H] and [Al/Fe]
are given in Table \ref{tabstars}.  For CS~29518--043 we had no blue
spectra and thus the aluminium abundance could not be computed.  For
homogeneity, in Table \ref{tabstars} the solar abundance of aluminium
has been taken from Grevesse and Sauval (\cite{GS00}) as in Cayrel et
al.  (\cite{CDS04}): $\rm log \epsilon_{(Al)\odot}=6.47$.  Note that
this value is very close to the value we obtained for the Sun, in the same
conditions as in the stars : ($\rm log \epsilon_{(Al)\odot}=6.43$).

\subsection{Abundance uncertainty}  \label{errab}
The typical uncertainty in the fit corresponds to 0.05~dex in abundance.
But the uncertainty due to random errors on the determination of the
stellar parameters must be added to this observational error.  The
random error in $T_{\rm eff}$ is close to 100K, the error in $\log g
\simeq 0.2$dex and the error in $\rm vt \simeq 0.2 km s^{-1}$.  In
Table~\ref{errors} $\rm \Delta [Al/Fe]$ is given for a typical model
of a giant and a turnoff star and for $\Delta T_{\rm eff}=100$K,
$\Delta \log g = 0.2$ and $\rm \Delta vt = 0.2km s^{-1}$.  If we
suppose that the errors are independent and we add them
quadratically, we find that the errors on [Al/Fe] due to the
uncertainties of the stellar parameters are close to 0.11 for the
giants and 0.07 for the dwarfs.  Taking into account the observational
error we find that the expected scatter is 0.12 for the giants and
0.08 for the turnoff stars.

\begin {table}[t]
\caption {[Al/Fe] uncertainties linked to stellar parameters.}
\label {errors}
\begin {center}
\begin {tabular}{lccc}
\hline
\multicolumn {4}{c}{GIANT STAR}\\
\multicolumn {2}{l}{CS~22948--066}\\
\multicolumn {4}{c}{A: $T_{\rm eff}$=5100K, log $g$=1.8 dex, vt=2.0 
km s$^{-1}$}\\
\multicolumn {4}{c}{B: $T_{\rm eff}$=5100K, log $g$=1.6 dex, vt=2.0 
km s$^{-1}$}\\
\multicolumn {4}{c}{C: $T_{\rm eff}$=5100K, log $g$=1.8 dex, vt=1.8 
km s$^{-1}$}\\
\multicolumn {4}{c}{D: $T_{\rm eff}$=5000K, log $g$=1.8 dex, vt=2.0 
km s$^{-1}$}\\
\hline
Abundance ratio   & $\Delta_{B-A} $ & $\Delta_{C-A} $& $\Delta_{D-A} $\\
\hline
[Fe/H]      &-0.04 & 0.02 &-0.05 \\
$[$Al I/Fe] & 0.06 & 0.05 &-0.04 \\
\hline
\multicolumn {4}{c}{TURNOFF STAR}\\
\multicolumn {2}{l}{CS~22177-09}\\
\multicolumn {4}{c}{A: $T_{\rm eff}$=6260K, log $g$=4.5 dex, vt=1.3 
km s$^{-1}$}\\
\multicolumn {4}{c}{B: $T_{\rm eff}$=6260K, log $g$=4.3 dex, vt=1.3 
km s$^{-1}$}\\
\multicolumn {4}{c}{C: $T_{\rm eff}$=6260K, log $g$=4.5 dex, vt=1.1 
km s$^{-1}$}\\
\multicolumn {4}{c}{D: $T_{\rm eff}$=6160K, log $g$=4.5 dex, vt=1.3 
km s$^{-1}$}\\
\hline
Abundance ratio   & $\Delta_{B-A} $ & $\Delta_{C-A} $& $\Delta_{D-A} $\\
\hline
[Fe/H]      &-0.02 & 0.03 &-0.05 \\
$[$Al I/Fe] & 0.02 &-0.03 &-0.03 \\
\hline
\end {tabular}
\end {center}
\end {table}

\begin{figure}
\resizebox  {8.4cm}{4.2cm}
{\includegraphics {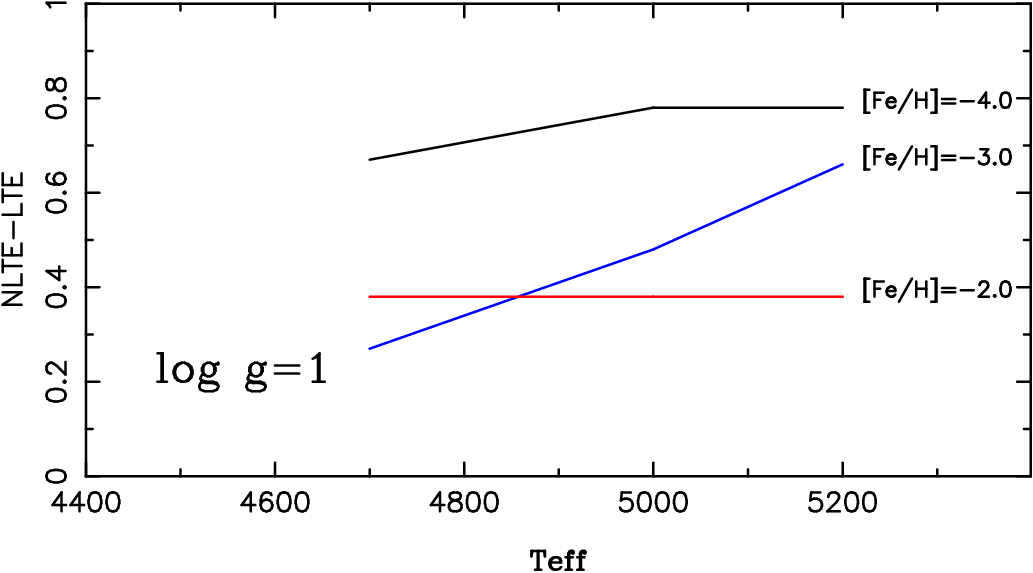} }
\resizebox  {8.4cm}{4.2cm}
{\includegraphics {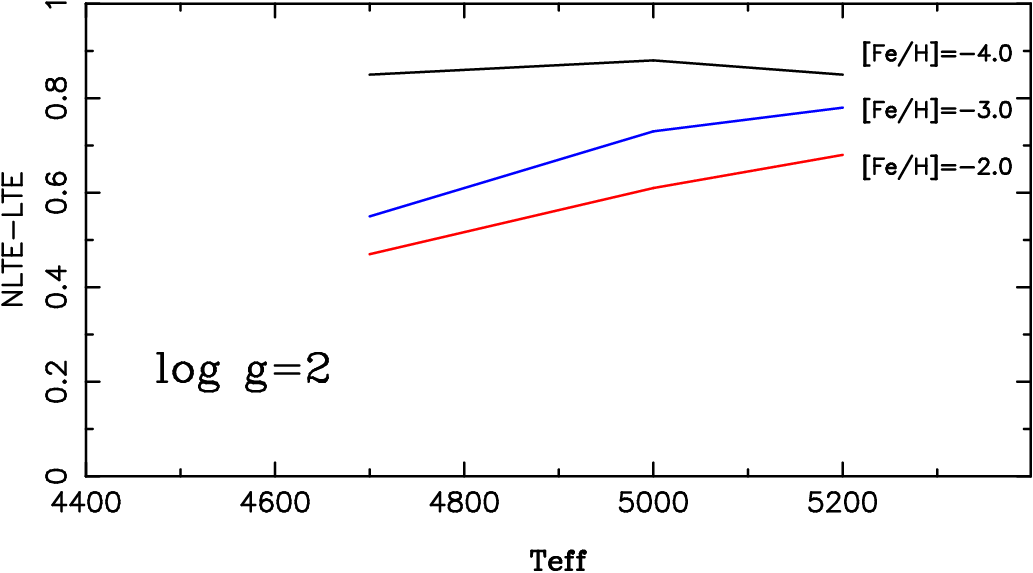} }
\resizebox  {8.4cm}{4.2cm}
{\includegraphics {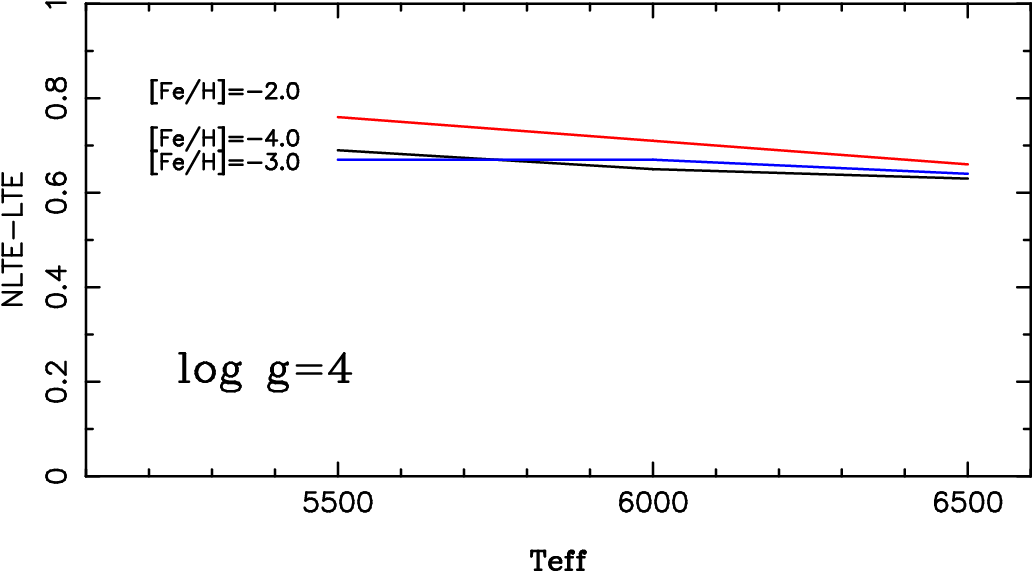} }
\caption[]{NLTE corrections behaviour with effective temperature, surface 
gravity and metallicity of the model}
\label{NLTEcor}
\end{figure}

\section{Results and discussion}
\subsection{[Al/Fe] in the early Galaxy} \label{algalax}

Figure \ref{NLTEcor} shows how the NLTE correction behaves with
effective temperature, surface gravity and metallicity of the model.
The correction depends on a balance between overionisation which
depopulates the lower levels (the photoionisation cross-section of
the ground state is exceptionally large) and a cascade of electrons
downward (phenomenon of "photon suction").  For aluminium,
photoionisation strongly dominates and the effect increases toward
higher $T_{\rm eff}$ and lower $\log g$.  Furthermore at low
metallicity the UV radiation field is stronger and hence the effect
also increases with decreasing [Fe/H] as  can be seen for the giants
($\log g=1$ and 2) in Fig.  \ref{NLTEcor}.  For turnoff stars,
according to Asplund (\cite{Asp05}), the correction reaches a maximum for
mildly metal-poor stars when the line is saturated.  This is also what
we observe in Fig.  \ref{NLTEcor} (lower panel).  Note that for
turnoff stars at $\rm [Fe/H] \approx -3$ the correction NLTE~--~LTE is
close to 0.6~dex, in good agreement with the computations of
Baum\"{u}ller \& Gehren (\cite{BG97}).

These corrections should not be used to determine a precise value of
the NLTE abundance of aluminium in metal-poor stars.  The shapes of
NLTE and LTE profiles are different (in particular when the resonance
lines of aluminium are strong) and a complete NLTE computation of the
profiles is necessary, as was done for the values given in Table  \ref{tabstars}.

\begin{figure}
\resizebox  {8.0cm}{4.5cm} 
{\includegraphics{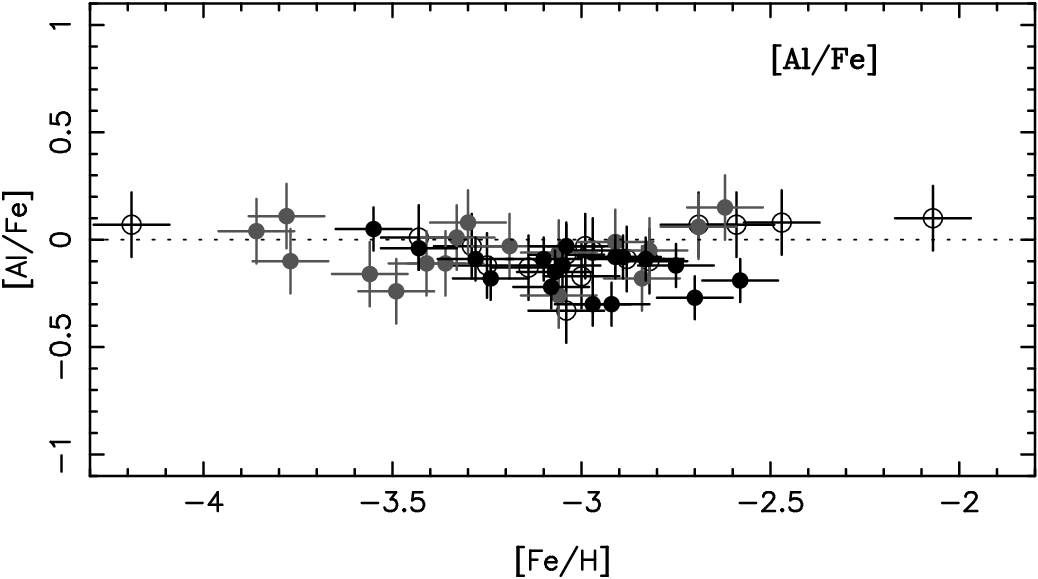}}
\caption {NLTE aluminium abundance in our sample of extremely 
metal-poor stars. The dark symbols represent the turnoff stars, the 
grey ones the unmixed giants and the open symbols the mixed giants.
}
\label{NLTE-Al}
\end{figure}

In Fig.  \ref{NLTE-Al} we have plotted [Al/Fe] vs.  [Fe/H] for our
samples of extremely metal-poor turnoff and giant stars.  There is a
good agreement between the abundances of aluminium in these two
different classes of stars.  Moreover the significant difference
between mixed and unmixed giants (see Fig. 8 in Spite et al.,
\cite{SCH06}) suggested by LTE determinations, is no longer observed.

In Andrievsky et al.  (\cite{ASK07}) it has been found that several
mixed giants of our sample were sodium-rich, suggesting that in these
stars, mixing is deep enough to bring to the surface the products of
the Ne-Na cycle.  Moreover it had been found (from LTE computations)
that all these Na-rich stars (but $\rm BD~17^\circ 3248$) seemed to be
also Al-rich. This peculiarity disappears with an NLTE
analysis of the aluminium lines.  In the Na-rich mixed giants, mixing
brings to the surface the products of the Ne-Na cycle but, as
expected, not the products of the Mg-Al cycle formed at higher
temperature.

At very low metallicity ($\rm [Fe/H]<-2.8$) in Fig.  \ref{NLTE-Al},
the relation [Al/Fe] versus [Fe/H] is rather flat, the global mean is
$\rm <[Al/Fe]>=-0.06 \pm 0.10$ (random error).  The scatter is very
close to the expected scatter when observational errors and
uncertainties on stellar parameters are taken into account (see
paragraph \ref {errab}).

Iron is generally used as a reference element in high resolution
spectral analysis because it has the largest number of usable lines
and thus the highest internal precision.  However the abundance ratio
[Al/Fe] in the ejecta of type II supernovae is very sensitive to the
mass cut.  Oxygen would be a better choice than iron, but in giants
the uncertainties on its abundance are large, and the oxygen abundance
could not be measured in the sample of dwarfs.  Magnesium may be a
good alternative (see Cayrel et al., \cite{CDS04}, Gehren et al.,
\cite{GLS04}, \cite{GSZ06}) since Mg and Al are both synthesised
during the advanced stages of the evolution of massive stars and since
the ratio [Al/Mg] is practically independent of the mass-cut.
Surprisingly, when we plot [Al/Mg] as a function of [Fe/H] (or [Mg/H])
the scatter at low metallicity is larger than the scatter of [Al/Fe]:
for $\rm [Fe/H]<-2.8$ $\rm <[Al/Mg]> = -0.32 \pm 0.14$ (random error).

Generally speaking for these EMP stars the scatter of [X/Mg] is, for
all the elements, larger than the scatter of [X/Fe] (Cayrel et al.,
\cite{CDS04}).  This increase of the scatter cannot be explained by
measurement errors on the magnesium abundance alone.  Following Gehren
et al.  (\cite{GSZ06}), NLTE significantly affects the determination
of the magnesium abundance, the maximum of the correction is about
0.15dex but it could be larger in giants.  This will be investigated
in a subsequent paper.

\begin{figure}
\resizebox{\hsize}{!} {\includegraphics{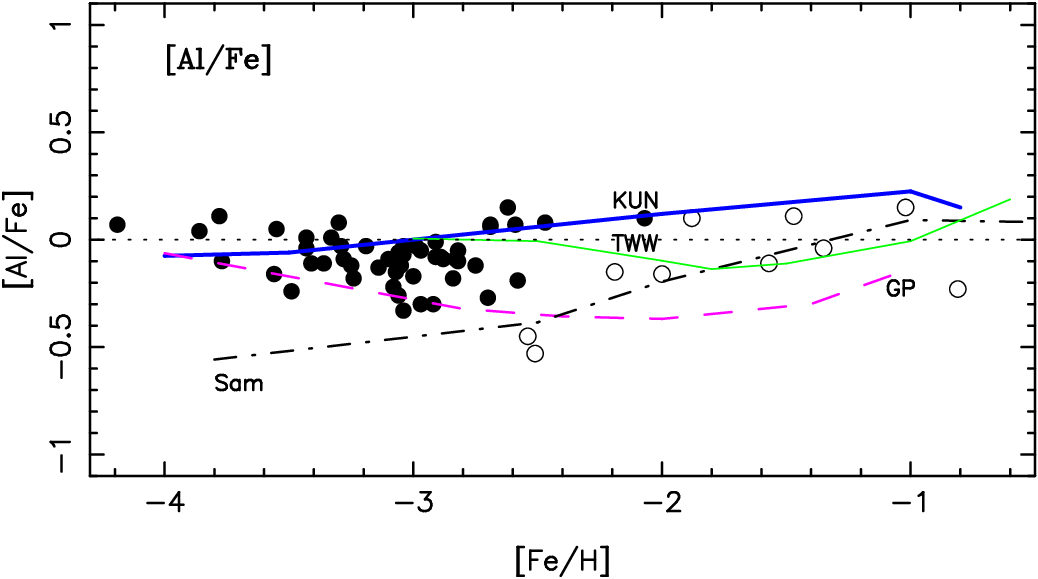}}
\caption[]{NLTE aluminium abundance in program stars vs.  their
metallicity.  The black dots represent our measurements and the open
circles the measurements of Gehren et al.  (2006) in halo stars.  The
theoretical predictions of Kobayashi et al., 2006 (KUN, full thick
line), Timmes et al., 1995 (TWW, full thin line, their curve ends up
at the metallicity [Fe/H]=--3), Samland, 1998, (Sam, dashed dotted
line), and Goswami \& Prantzos, 2000 (GP, dashed line) are also
represented.  }
\label{NLTEab}
\end{figure}

\subsection{[Al/Fe] and the models of the chemical evolution of the Galaxy}

The scatter (0.12~dex around the mean value) of the NLTE values of
[Al/Fe] is smaller than the scatter of the LTE determinations and can
be explained by measurement errors only.  As a consequence, the value
of the yields of aluminium in massive supernovae, deduced by Tsujimoto
et al.  (\cite{TSY99}) from the "large scatter" of [Al/Fe] at low
metallicity, can be questioned.

In Fig.  \ref{NLTEab} we compare our measurements of [Al/Fe] in EMP
stars to the measurements of Gehren et al.  (\cite{GSZ06}) in a sample
of nearby halo dwarfs.  These measurements extend our sample to
higher metallicities.  There is a rather good agreement between the
two sets of measurements: the mean [Al/Fe] value deduced from the halo
stars of the sample of Gehren et al.  is $\rm [Al/Fe]=-0.13 \pm 0.2$
~(vs.  [Al/Fe]~=~--0.08, section \ref{algalax}).

However the two most metal-poor stars of the sample of Gehren et al.
(\cite{GSZ06}) seem to be aluminium-poor relative to our sample.
These two stars are among the faintest objects observed by Gehren et
al.  and thus (see their section 2) the S/N of the spectra is probably
as low as 100 near $\rm H{\alpha}$ and even lower in the region of the
aluminium lines.  As a consequence for these stars the abundance error
should be higher than the average abundance error quoted by Gehren et
al.: 0.10~dex.  One of these stars (G~48-29 = HE~0938-0114) has been
also observed by Barklem et al.  (\cite{BCB05}) with UVES at the VLT.
The S/N of their spectrum is higher than 300 and they found an LTE
ratio of [Al/Fe]~=~--0.78.  If we apply to this value a NLTE-LTE
correction deduced from the computations of Gehren et al.  (2006,
their Table  2), we find [Al/Fe]~=~--0.17, in good agreement with our
measurements at this metallicity.

It is interesting to compare the evolution of [Al/Fe] in the Galactic
halo with the predictions of the different models of the Galactic
evolution based on the element yields of massive stars (Fig.
\ref{NLTEab}).  In the region $\rm -4<[Fe/H]<-2.5$ the predictions of
Samland (1998) and of Goswami \& Prantzos (2000) for the ratio [Al/Fe]
are too low compared to the observations.  A better agreement is
obtained when the measurements are compared to the predictions of
Timmes et al.  (1995) or Kobayashi et al.  (2006).  Timmes et al.
(1995) calculated the evolution of the aluminium-to-iron ratio down to
[Fe/H]=--3 based on the  SNe~II metallicity-dependent
yields of Woosley \& Weaver (1995).
The best agreement is found by assuming that the
iron production by SNe~II should be decreased by a factor of 2
or slightly less (see Fig.  19 in Timmes et al.)  as is generally
found (see also Andrievsky et al., 2007).  
Our data are in fair agreement with
the [Al/Fe] ratios predicted by the Galactic chemical evolution model of  
Kobayashi et al. (2006).
The yields used  by these authors are calculated for very low
metallicity and the explosion energies of the supernovae/hypernovae
are based on the fitting of the  light curve and 
the spectra of individual supernovae/hypernovae.  

Recently Tominaga et al.  (\cite{TUN07}) have computed the predictions
of inhomogeneous models of the Galactic enrichment.  In these models
the observed trends of [X/Fe] vs.  [Fe/H] are not due to metallicity
effects but to the combination of the progenitor (hypernovae,
supernovae) masses and explosion energies.  For most of the elements
the predicted slope is weak and compatible with the observations but a
strong slope of the ratios [Na/Fe] and [Al/Fe] vs.  [Fe/H] is expected
since the hypernovae with high explosion energies, that are supposed to
explode at the very beginning of the Galaxy, eject much less sodium and
aluminium than the regular supernovae that explode a little later.
Our observations show no significant slope of [Na/Fe]
(Andrievsky et al.  \cite{ASK07}) nor of [Al/Fe].  However these
observations can be reconciled with the computations of Tominaga et
al.  if we suppose that in the interval $\rm -4<[Fe/H]<-2.5$ the
galactic gas was efficiently mixed.  They predict that in the
ejecta of their "mean" supernova/hypernova (integrated over the IMF)
[Na/Fe] $\simeq$ [Al/Fe] $\simeq$ --0.3.  These values correspond
rather well to the observed values: $\rm [Na/Fe] =-0.21 \pm 0.13$
(Andrievsky et al., \cite{ASK07}) and $\rm [Al/Fe]=-0.08 \pm 0.12$.

\section{Conclusion}
The NLTE abundances of Al computed in this work for very metal-poor
stars provide more reliable values of the ratio Al/Fe and a more
reliable trend of this ratio versus metallicity.  The scatter is
significantly reduced compared to what it was for the LTE
determinations (see Fig.  8 of Cayrel et al.  2004 for the giants);
the mean value of the Al/Fe ratio at the lowest metallicity is solar:
$\rm [Al/Fe] = -0.06 \pm 0.10$, the error quoted here is the $1
\sigma$ value of the scatter around the mean regression line.  This
small scatter can be explained entirely by the measurement error and
the random error on the stellar parameters.

The Al abundance is the same in extremely metal poor TO stars, "mixed"
and "unmixed" giants.  Moreover, some "mixed" giants that were found
to be Na-rich in Andrievsky et al.  (2007) were then suspected to be also
Al-rich.  In fact this apparent enrichment of aluminium in these
Na-rich giants is entirely explained by NLTE effects.  After a correct
treatment of the aluminium lines, the Na-rich giants have a normal
[Al/Fe] ratio.

The trend, now clearly defined, has a very small slope: owing to
possible systematic errors, it could well be zero.  In the region $\rm
-4<[Fe/H]<-2.5$ the behaviour of [Al/Fe] may thus be easily compared
with the predictions of the models of chemical evolution, favouring:\\
--the predictions of Timmes et al. (\cite{TWW95}), (but their 
predictions end at $\rm [Fe/H]=-3$), or\\
--the predictions of Kobayashi et al. (\cite{KUN06}).

Our observations can be reconciled with the predictions of Tominaga et
al.  (\cite{TUN07}) if we assume an efficient mixing of the ejecta of
the hypernovae - supernovae in the early Galaxy for metallicities
above $\rm [Fe/H]=-4$.

The NLTE Na computations in extremely metal-poor stars provide, for
the unmixed and not Na-enriched stars, a well defined mean value of
$\rm [Na/Fe] = -0.21 \pm 0.13$ (Andrievsky et al., 2007).  Now the
NLTE Al abundances provide reliable Al/Na ratios: [Al/Na] = 0.15~dex.
This value is slightly higher than solar, but owing to the error bars,
it could be solar.

Spite et al. (\cite{SCP05}), commenting
on the wide scatter in nitrogen abundances found among EMP stars,
noted that rotating massive stars
could possibly influence the abundances of the first stars.  These
rotating stars (Meynet et al.  2005, 2006) produce powerful winds that
may be slightly rich in Na and Al.  The ratio of the mass fractions of
Al and Na in such winds, computed by Decressin et al.  (\cite{DMC07}),
appears to be quite different for different models and phases.  Thus
the observed value of the Al/Na ratio, although compatible with the
production of such winds, does not imply that this production amounts
to a significant contribution to the abundances of EMP stars.  On the
other hand, the Al/Na ratio is incompatible with the predictions of
Meynet et al.  (2005, 2006) for the yields of rotating $7M_{\odot}$
E-AGB stars that can explain the abundances in the carbon-rich
extremely metal-poor stars, leaving a question mark about the
associated primary production of nitrogen.

In spite of significant progress, this work suggests that the quest
for even more reliable constraints on the yields of the first
supernovae would require the interpretation of the measurements by
sophisticated models including not only the NLTE corrections, but also
computation of the diffusion of metals inside the atmosphere, use of
(time consuming) 3D models, as well as better physical constants
(collisions, oscillator strengths, molecular constants...).

\begin {acknowledgements} This work has been supported by the
"Programme National de Physique Stellaire" (CNRS).  S.A. kindly
acknowledges the Paris-Meudon Observatory for its financial support
during his stay in Meudon and the laboratory GEPI for its hospitality.
\end {acknowledgements}
\balance

\end{document}